\documentclass[11pt]{article}
\usepackage{moriond,epsfig}
\usepackage{epstopdf}

\def\be{\begin{equation}}
\def\ee{\end{equation}}
\def\bea{\begin{eqnarray}}
\def\eea{\end{eqnarray}}

\begin{document}
\vspace*{4cm}
\title{STUDY OF VECTOR BOSON FUSION HIGGS IN ATLAS-LHC}

\author{ D. VAROUCHAS }

\address{LAL, Univ Paris-Sud, IN2P3/CNRS, Orsay, France}

\maketitle\abstracts{
Within the framework of Standard Model, the production mode of Higgs boson through the fusion of the vector bosons $W$ or $Z$ ({\it Vector Boson Fusion}) is one of the most important production mechanisms, providing a specific detection signature. A detailed study regarding this issue is being undergone for ATLAS detector in LHC and some general features of this analysis are being presented in this note emphasizing in the study of Central Jet Veto.}

\section{VBF Topology and its characteristics}
\begin{figure}
\centering
\begin{tabular}{cc}
\includegraphics[width=6cm]{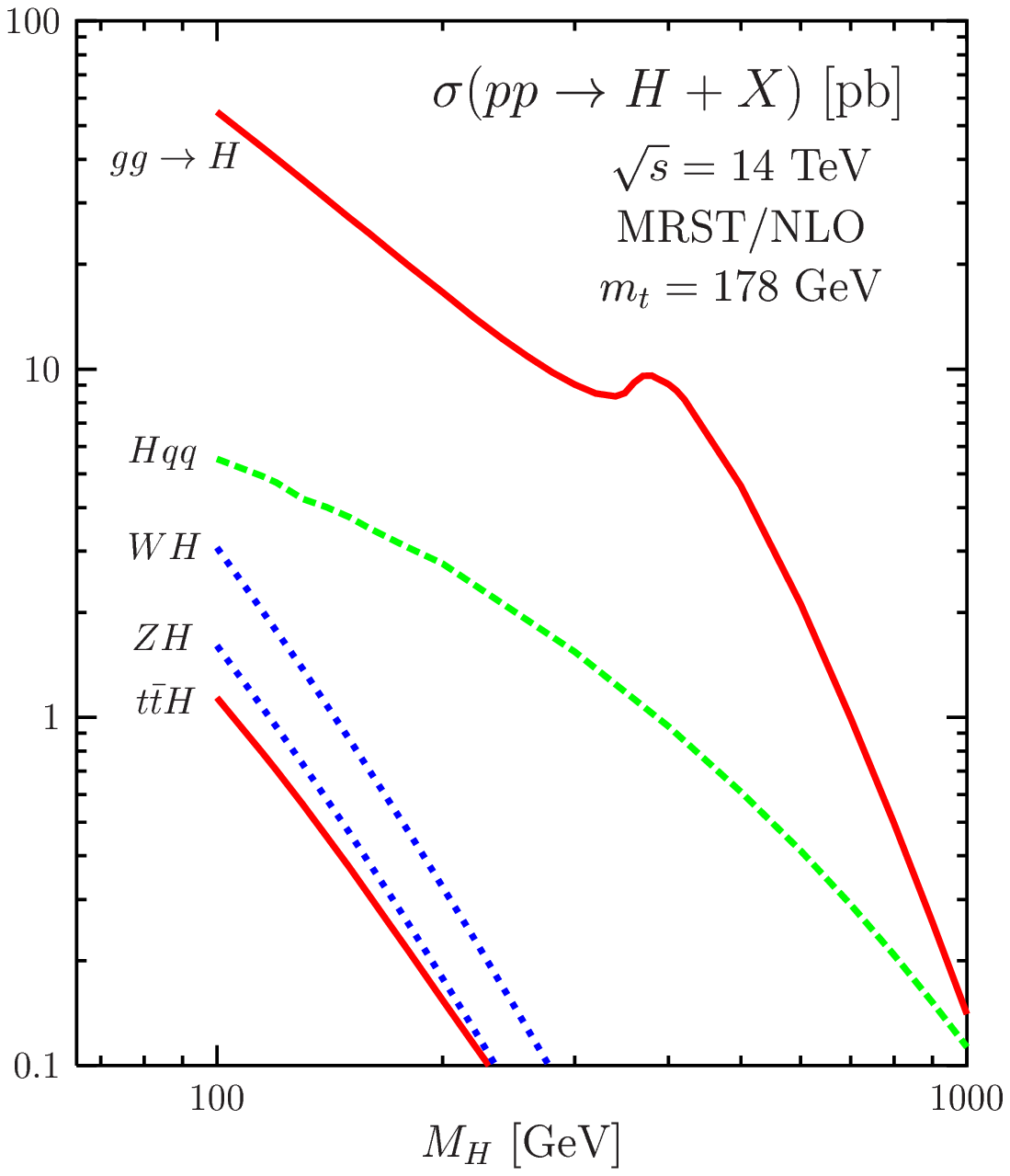} &
\includegraphics[width=6cm]{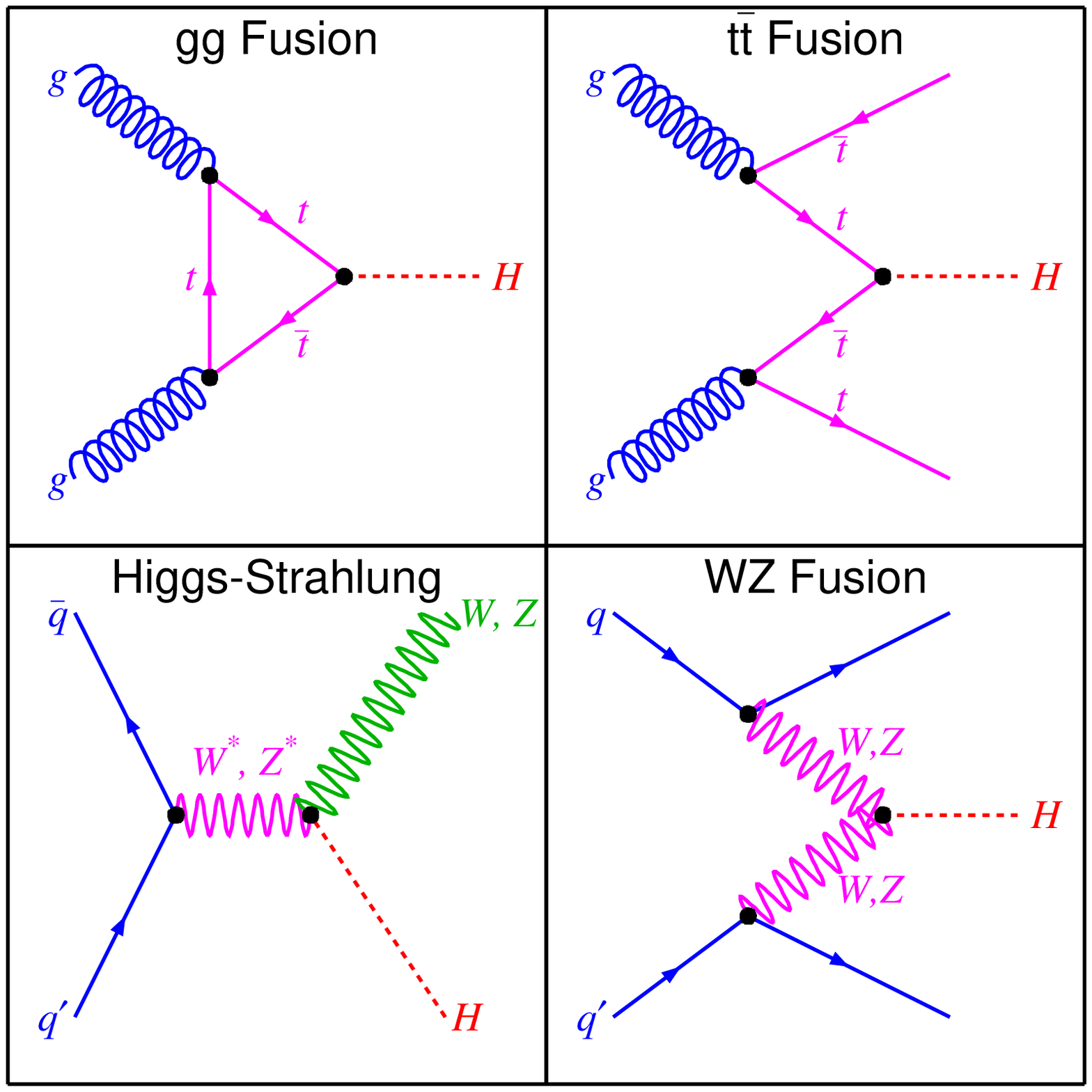}  
  \end{tabular}
\caption{ \label{pp} (a) Cross Section of Higgs boson  for (b) the four different production mechanisms  in LHC, within the frame of Standard Model. }
\end{figure}

The four different production mechanisms of a Standard Model Higgs boson \cite{djouadi} at the LHC are (Fig.\ref{pp}b): the $gg$ fusion, the Vector Boson Fusion, the Higgs boson production associated with vector bosons $W$ or $Z$ and the Higgs boson production associated with the production of a $t\overline{t}$ pair. Although VBF cross section is one order of magnitude lower than the dominant one of $gg$ fusion (Fig.\ref{pp}a), its topological characteristics provide a signature which makes it an important discovery channel for a low mass Higgs boson.

As it is illustrated   in Fig.\ref{pp}b the VBF\footnote{Vector Boson Fusion} Higgs (or $WZ$ Fusion) occurs when two quarks originating from the proton beam are being scattered through the exchange of a $W$ or $Z$ boson and their fusion produces the Higgs boson. The important aspect of this topology is the fact the two scattered quarks give two energetic jets highly separated in rapidity in the forward regions of the detector. 
  
Because of the electroweak character of the VBF Higgs boson production ($W,Z$ colorless exchange), a low gluon radiation activity is expected which is translated in detection terms as a low jet activity. Consequently, no jets are expected in the central region of the detector above a certain $p_{T}$ limit, in addition to the Higgs boson decay products. This aspect is motivating a central jet veto: the event is rejected if a third jet is detected in the central region above a given threshold of $p_{T}$. This cut is efficient against the main backgrounds and especially against $t\overline{t}$ and can be a useful tool to the improvement of the S Vs B separation.

\section{ VBF $H \rightarrow \tau^{+} \tau^{-}$ in ATLAS} 
The most promissing part of the VBF Higgs analysis is the one which studies its decay into a pair of $\tau^{+} \tau^{-}$ leptons. This channel was first studied in ref.\cite{Asai} for ATLAS \cite{TDR} and now, a more detailed study is being undergone using state of the art Monte Carlo generators and a fully detailed simulation of the detector. 

Due to the weak branching ratio of Higgs boson decaying to $\tau^{+} \tau^{-}$ for masses $m_{h}>140 GeV/c^{2}$, the analysis is performed in the mass range of $110GeV/c^{2}<m_{h}<140 GeV/c^{2}$ and it is divided into three different categories depending on the decays of $\tau$ leptons (leptonic or hadronic decay). As a consequence, three sub-channels are obtained at the end: The lepton-lepton channel with a B.R. $\simeq$ 12 \%, the lepton-hadron channel with a B.R. $\simeq$ 46 \% and the hadron-hadron channel with a B.R. $\simeq$ 42 \%.

For the processes of background the generated events correspond to: 
\begin{list}{\textbullet}{}
\item $Z \rightarrow \tau^{+} \tau^{-} + jets$ 
\item $t \overline{t}$ 
\item $W^{\pm}  + jets$ 
\end{list}
Among them the dominant background is the $Z \rightarrow \tau^{+} \tau^{-} + jets$ which has a quite similar topology with the signal and therefore is the most difficult to suppress. 

The final goal is to reconstruct the invariant mass of the couple $\tau^{+} \tau^{-}$ which equals to the mass of Higgs boson. The existence of at least two neutrinos in the event, requires the use of a collinear approximation to estimate the $m_{\tau ^{+} \tau^{-}}$ invariant mass from the measure of the missing $p_{T}$ vector and the $p_{T}$ of the visible tau decay products. This approximation implies that the tau decay products are collinear to the direction of taus. This is an acceptable assumption considering the hig Pt of the $\tau$ leptons. However it provides a good mass resolution only if the tau decay products are not back to back, which is the case if the parent Higgs boson has reasonably high Pt.   

The first step of the analysis consists of the identification of leptons and hadrons following the general ATLAS methods. Then, several cuts are applied in order to separate background from signal. This cuts can be categorized in three groups. The first group is related to the tau decay and different constraints concerning the kinematic variables of leptons or jets originating from the hadronic decay of tau are being applied. These cuts differ in lepton-lepton, lepton-hadron and hadron-hadron cases. Next, the following set of cuts is based on collinear approximation method used for the mass reconstruction and finally, several constraints deriving from the special jet VBF behavior described in the previous paragraph are being applied such as, a given $p_{T}$ threshold for the two forward jets, a large separation in $\eta$, the decay products to lie between the forward jets in $\eta$, high invariant mass of the two forward jets and also a central jet veto.

\begin{figure}
\centering
\includegraphics[width=6cm]{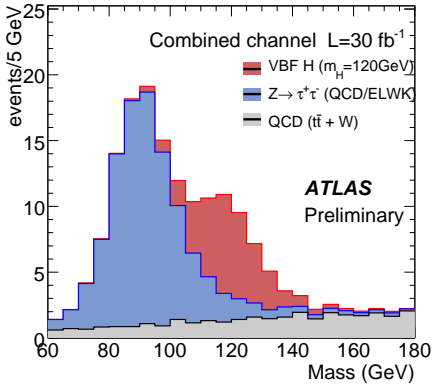} 
\caption{ \label{Higgs_m} The reconstructed $\tau^{+}\tau^{-}$ invariant mass for a VBF Higgs boson signal of $120GeV/c^{2}$ in the combined lepton-lepton and lepton-hadron channel above all backgrounds after application of all cuts. The number of signal and background events is shown for an integrated luminosity of $30fb^{-1}$.  }
\end{figure}

At Fig.\ref{Higgs_m} we see the mass distribution after having applied all the different cuts. It is obvious that the $Z \rightarrow \tau^{+} \tau^{-}$ is the dominant background as expected, but still there is a clear peak related to the mass of Higgs boson. Nevertheless, the importance of the mass resolution is highlighted through this plot, which shows that a worse resolution would result to a contamination of the signal by the $Z$ resonance. 

\section{Central Jet Veto}
\begin{figure}
\centering
\includegraphics[width = 11cm]{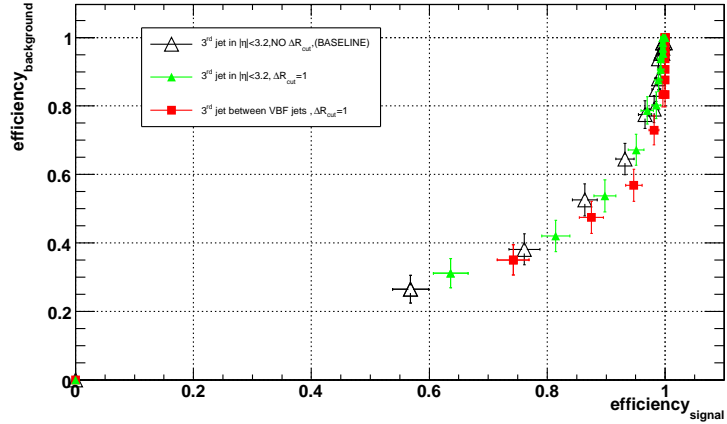} 
\caption{ \label{perf} Jet veto cut efficiency for signal and background. The white triangles represent the primary jet veto method where no $\Delta R$ cut is applied and the third jet is searched in $|\eta|<3.2$ . The green triangles represent the primary jet veto method but searching the third jet only between the two forward jets in term of $\eta$ and finally, the red squares represent the previous jet veto method having applied the $\Delta R$ cut as well. Every point in all curves corresponds to the efficiencies of a given $p_{T}$ threshold of the third jet starting from $10 GeV/c$ with a step of $5 GeV/c$.  }
\end{figure}

Jets play an important role in this channel and therefore, many studies have been done towards this issue. In this analysis, the jets are reconstructed with a cone algorithm of a radius of 0.4 and part of these jet studies related to VBF Higgs topology, is the optimization of the central jet veto cut. This cut rejects an event if a third jet is found in a region of $|\eta|<3.2$ having a $p_{T}>20 GeV/c$. Other alternatives of this cut are also studied by changing this $|\eta|$ region. A jet veto was studied without applying any constraints on $\eta$ and also applying the cut if a third jet is only found between the two forward jets in $\eta$. Another interesting feature seen, was that in many cases of signal events rejected by this cut, the vetoing jet was very close in $\Delta R$  \footnote{$   \Delta R=\sqrt{\Delta \eta^2 + \Delta \phi^2}$} to one of the two forward jets. This was due to the splitting of the initial jet originating from the scattered parton. Using a larger jet reconstruction cone, such as 0.7 for instance which is the other alternative in ATLAS algorithms, the jet splitting effect might have been reduced but the overall signal significance would still be lower. Therefore, a better option is to introduce an additional constraint, which is to not take into consideration the third jet if it is close to one of the forward jets, and more precisely, if $\Delta R$ between these two jets is less than 1. In order to compare the three different approaches the cut efficiency is calculated for Signal and Background. The results are shown in Fig.\ref{perf}.

 With the combination of the $\Delta R$ condition to eliminate the vetoing jets close to the forward jets and searching for a third jet between the two forwards jets in terms of $\eta$, a gain of $\sim$10\% for the signal jet veto cut efficiency is achieved, as it is seen in Fig.\ref{perf}. 

\section{Outlook}
The main characteristics of VBF Higgs boson topology and some general aspects of the analysis of its decay in a couple of $\tau^{+} \tau^{-}$ in ATLAS were presented. An example of improving a cut efficiency was addressed and it was shown that a gain of 10 \% is accomplished modifying the jet veto cut. This analysis is not yet finalized since many studies are still being undergone, in particular concerning the effect of $pile-up$, which causes the creation of more jets into both signal and background events for the high LHC luminosities. Nevertheless, it was shown that VBF Higgs boson is one of the most promising channels for a low mass Higgs boson.

\end{document}